\documentclass[a4paper,fleqn]{cas-sc}
\ExplSyntaxOn
\exp_args:NNno \exp_args:Nno \use:n { \cs_gset:Npn \__make_fig_caption:nn #1#2 }
  {
    \exp_after:wN \use_ii_i:nn \exp_after:wN
      { \__make_fig_caption:nn {#1} {#2} }
      { \dim_set:Nn \l_fig_width_dim \linewidth }
  }
\exp_args:NNno \exp_args:Nno \use:n { \cs_gset:Npn \__make_tbl_caption:nn #1#2 }
  {
    \exp_after:wN \use_ii_i:nn \exp_after:wN
      { \__make_tbl_caption:nn {#1} {#2} }
      { \dim_set:Nn \l_tbl_width_dim \linewidth }
  }
\ExplSyntaxOff 
\usepackage{duckuments}
\usepackage[numbers]{natbib}
\usepackage{mathtools}
\usepackage{amsmath}
\usepackage{graphicx}
\usepackage{subcaption}
\usepackage{rotating} 
\usepackage{caption}
\usepackage{subcaption}
\usepackage{commath}
\def\tsc#1{\csdef{#1}{\textsc{\lowercase{#1}}\xspace}}
\tsc{WGM}
\tsc{QE}
\tsc{EP}
\tsc{PMS}
\tsc{BEC}
\tsc{DE}

\begin{document}
\let\WriteBookmarks\relax
\def\floatpagepagefraction{1}
\def\textpagefraction{.001}

\shorttitle{Gravitational Lensing in Loop Quantum Cosmology}

\shortauthors{Rownak Kundu et~al.}

\title [mode = title]{Gravitational Lensing of Dark Energy Models and $\Lambda$CDM Using Observational data in Loop Quantum Cosmology}                      
\author[1,2]{Rownak Kundu}
\ead{rownakkundu@gmail.com}
\address[1]{Department of Mathematics, Nalbari College, Nalbari-781335, Assam, India.}
\address[2]{Department of Mathematics, Indian Institute of
Engineering Science and Technology, Shibpur, Howrah-711 103,
India.}
\author[2]{Ujjal Debnath}
\ead{ujjaldebnath@gmail.com}
\author[4,5,6]{Himanshu Chaudhary}
\cormark[1]
\ead{ujjaldebnath@gmail.com}
\address[4]{Department of Applied Mathematics, Delhi Technological University, Delhi-110042, India}
\address[5]{Pacif Institute of Cosmology and Selfology
(PICS), Sagara, Sambalpur 768224, Odisha, India}
\address[6]{Department of Mathematics, Shyamlal College,
University of Delhi, Delhi-110032, India.}
\author[7]{G. Mustafa}
\cormark[1]
\ead{gmustafa3828@gmail.com}
\address[7]{Department of Physics, Zhejiang Normal University, Jinhua 321004, People’s Republic of
China}
\cortext[cor1]{Corresponding author}
\begin{abstract}
This paper investigates the accelerated cosmic expansion in the late Universe by examining two dark energy models, viscous modified Chaplygin gas (VsMCG) and variable modified Chaplygin gas (VMCG), within loop quantum cosmology alongside the $\Lambda$CDM model. The objective is to constrain cosmic parameters using the $\Lambda$CDM model and 30 of the latest $H(z)$ measurements from cosmic chronometers (CC), including Type Ia Supernovae, Gamma-Ray Bursts (GRB), Quasars, and 24 uncorrelated baryon acoustic oscillations (BAO) measurements across a redshift range from 0.106 to 2.33. The latest Hubble constant measurement from Riess in 2022 is included to enhance constraints. In the $\Lambda$CDM, VsMCG, and VMCG frameworks, best-fit parameters for the Hubble parameter ($H_0$) and sound horizon ($r_d$) are obtained. The results highlight significant disparities between $H_0$ and $r_d$ values from late-time observational measurements, reflecting the known $H_0$ and $r_d$ tensions. The gravitational lensing optical depth of the two dark energy models is studied by plotting $\log(\tau(z_l)/H_0^{-3} \tau_N)$ vs $z_l$. The probability of finding gravitational lenses (optical depth) in both models increases with lens redshift $z_l$. The change in optical depth behavior for different parameter constraints is graphically analyzed. A joint analysis of VsMCG and VMCG with $\Lambda$CDM is conducted. While the models diverge in the early Universe, they are indistinguishable at low redshift. Using the Akaike information criteria, the analysis indicates that neither dark energy model can be dismissed based on the latest observations.
\end{abstract}

\maketitle
\section{Introduction}
Our universe is undergoing accelerated expansion, a phenomenon supported by multiple independent studies \cite{1,2,3,4} conducted over the past two decades. While the actual cause of this accelerated phenomenon is still debatable, many attribute it to huge negative pressure build-up in our present Universe. Many assume that some kind of `mysterious’ energy must be at play here and have named it `dark energy’ (DE, hereafter). Dark energy, or DE, almost occupies $70\%$ of our Universe, so its study has been of immense importance. Over the years, researchers have proposed numerous such DE candidates to explain our accelerating Universe, and among them, the cosmological constant $\Lambda$ happens to be the simplest. The other candidates includes Chaplygin gas \cite{5,6}, quintessence \cite{7,8}, phantom energy \cite{9}, holographic dark energy \cite{10,11}, among the numerous others in \cite{12}. Despite the diverse range of proposed dark energy candidates, the key to unravelling the mysteries of dark energy lies not only in theoretical constructs but also in empirical observations. Observational cosmology has thus become instrumental in discerning the nature of dark energy. Recent findings from the DESI collaboration suggest that the $\Lambda$CDM model, which describes the Universe's expansion in terms of dark energy and cold dark matter, may require reevaluation. DESI Luminous Red Galaxy data at $z_{\text{eff}} = 0.51$ shows a significant discrepancy in $\Omega_{m0}$ values, challenging the Planck-$\Lambda$CDM cosmology \cite{13}. \cite{14} suggested that resolving $\Lambda$CDM tensions requires redshift and scale consistency checks to identify missing physics, addressing $H_0$ and $S_8$ discrepancies. Moreover, a similar analysis could be performed by considering the look-back time and cross-correlations, potentially resolving tensions in the data, as discussed in \cite{15,16,17}.\\\\
Among the observational probes employed by cosmologists, the study of the large-scale structure of the Universe stands out prominently. Observations of the large-scale structure, which include the spatial distribution of galaxies, galaxy clusters, and cosmic voids, offer invaluable insights into the underlying dynamics of cosmic evolution \cite{18}. By scrutinizing the cosmic web on vast scales, cosmologists can glean crucial information about the composition of the Universe, its expansion history, and the influence of dark energy on cosmic structures \cite{19}. However, as observational techniques have advanced, they have revealed a perplexing tension between early-time and late-time cosmological observations. This tension manifests in discrepancies between measurements obtained from the early Universe, such as those derived from the cosmic microwave background (CMB) radiation \cite{20} and those obtained from more recent observations of the local Universe, such as those from Type Ia supernovae \cite{21} and Baryon Acoustic Oscillations (BAO). BAO plays a pivotal role, offering a unique and powerful tool for understanding the large-scale structure of the Universe. These oscillations are imprints left on the distribution of matter in the early Universe, originating from the coupled interaction between photons and baryons (protons and neutrons) before the era of recombination \cite{22}. The significance of BAO lies in their ability to serve as standard rulers on cosmological scales. As the Universe expands, these primordial density fluctuations lead to characteristic patterns in the distribution of galaxies and cosmic microwave background radiation. The formation of BAO can be traced back to the sound waves that propagated through the primordial plasma of the early Universe. As the Universe expanded, the sound waves left behind a distinctive pattern of overdense and underdense regions, creating a preferred scale known as the BAO scale or sound horizon \cite{23}. This scale acts as a standard ruler, allowing cosmologists to measure the geometry and expansion rate of the Universe over cosmic time. The BAO scale, or sound horizon, serves as a crucial cosmological standard because its size is determined by fundamental cosmological parameters, such as the density of baryons and dark matter, as well as the speed of sound in the primordial plasma. Observations of BAO in the large-scale distribution of galaxies provide a powerful constraint on these parameters and offer insights into the nature of dark energy \cite{24}. In the field of cosmology, the study of BAO and the sound horizon has become an essential tool for precision cosmological measurements. Large galaxy surveys, such as the Sloan Digital Sky Survey (SDSS), have played a key role in mapping the distribution of galaxies and detecting the BAO signal \cite{25}. By analyzing the BAO scale in the clustering of galaxies, researchers can infer the expansion history of the Universe and shed light on the mysterious components of dark matter and dark energy. Thus, Baryon Acoustic Oscillations stand as a cornerstone in our quest to unravel the fundamental properties and evolution of the cosmos.\\\\
While BAO provide valuable insights about the large-scale structure of the Universe and the behavior of dark energy, the accelerating expansion of the cosmos remains a profound puzzle that has spurred researchers to explore alternative explanations. Many assume that Einstein's gravitational theory might be incomplete. So, researchers modified the underlying gravity theory (and hence, the name modified theory of gravity \cite{26,27}) to incorporate the accelerating phase of our Universe. While these theories effectively explain the formation of structures and gravitational interactions on a large scale, quantum gravity becomes necessary to understand behaviors at smaller scales. Further, the backward evolution of our Universe in time results in the collapse of our Universe into a single point with diverging energy density. Such instances lead to the failure of our classical gravity theory, rendering it ineffective in describing the unfolding events. Quantum gravity, with its distinct dynamics on smaller scales, is expected to solve this dilemma. One such theory based upon quantum gravity is loop quantum cosmology (LQC, hereafter) \cite{28,29}. In recent years, several DE models have been studied in its framework. Jamil et al. \cite{30} investigated the combination of modified Chaplygin gas with dark matter within the LQC framework, resolving the cosmic coincidence problem. The authors in \cite{31}, discovered that loop quantum effects could prevent future singularities in the FRW cosmology. Given its importance, in our present study as well, we considered two models of Chaplygin gas, namely, viscous modified Chaplygin gas (VsMCG) and variable modified Chaplygin gas (VMCG) in LQC framework, to analyze their optical depth behaviour by constraining the model parameters to their best-fit values. Further, the significance of the study of gravitational lensing also has long been recognised in cosmology, and the phenomenal work of Refsdal et al. \cite{32} in this context is worth mentioning. Over recent years, gravitational lensing emerged as an important tool for studying dark matter, dark energy, and black holes, among many others. Gravitational lensing optical depth represents the likelihood of the formation of multiple images owing to the influence of gravitational lenses in our Universe. It was first used by. \cite{33,34} and consequently by many others \cite{35,36,37}, including Kundu et al. \cite{38,39} to study and understand dark energy models. Our present paper uses this lensing phenomenon to qualitatively analyse the two DE models against their constraint parameter values and record the changes graphically. This methodology of studying the gravitational lensing optical depth of DE models can give us an estimation in understanding the large-scale structure of our Universe, thereby helping us better understand our models.\\\\
Finally, our paper is organized as follows: Section \ref{lqc} delves into the background equations of the two dark energy models within the framework of loop quantum cosmology. Section \ref{methodology} outlines the methodology employed to constrain the parameters of the dark energy models using various datasets. Section \ref{lensing} is dedicated to the study of the lensing phenomenon, including the derivation of the equation for optical depth to be used in our study. In Section \ref{results}, we present the outcomes of our study, and we conclude by discussing our findings in Section \ref{conclusion}.
\section{Background equations: Loop Quantum Cosmology} \label{lqc}
In loop quantum cosmology, for a isotropic and homogeneous
Universe the background equations are defined by:
\begin{equation}\label{frw-1}
    H^2= \frac{1}{3}\left(1-\frac{\rho}{\rho_1}\right)\rho,
\end{equation}
\begin{equation}\label{frw-2}
    \dot H =-\frac{1}{2}\left(1-\frac{2\rho}{\rho_1}\right)(\rho+ p)
\end{equation}
where $\rho_1=\frac{\sqrt{3}}{16 \pi^2 \gamma^3 G^2 \hbar}$ is the
critical loop quantum density and $\gamma$ the dimensionless
``Barbero-Immirzi parameter''. The conservation equation in LQC is
given by:

\begin{equation}\label{continuity-eqn}
    \dot \rho +3H(\rho +p)=0
\end{equation}
Assuming the matter content of our Universe to be a combination of
DM and DE, the quantities $\rho$ and $p$ defined in Eqs
(\ref{frw-1}) and (\ref{frw-2}) will modify to $\rho= \rho_m+
\rho_d$ and $p= p_m +p_d$. With the consideration that both DM and DE
are separately conserved, the equation
(\ref{continuity-eqn}) now seperates into two distinct equations
as follows:

\begin{equation}\label{dm}
     \dot \rho_m +3H(\rho_m +p_m)=0,
\end{equation}
\begin{equation}\label{de}
     \dot \rho_d +3H(\rho_d +p_d)=0,
\end{equation}
Since, pressure in dark matter is very negligible (i.e. $p_m=0$),
equation (\ref{dm}) gives $\rho_m= \rho_{m0}(1+z)^3$. Here,
$\rho_{m0}$ represent the density value at the present epoch.
\subsection{Viscous modified Chaplygin gas (VsMCG)}
The expression for pressure in the case of viscous modified
Chaplygin gas (VsMCG) is given by \cite{40}:
\begin{equation} \label{pressure_VMCG}
    p_d= A\rho_d-\frac{B}{\rho_d^\alpha}- 3\zeta_0 \sqrt{\rho_d}H
\end{equation}
where $A$, $B$, $\alpha$ and $\zeta_0$ are constants. Substituting $p_d$ from
Eq (\ref{pressure_VMCG}) in Eq (\ref{de}), we get:
\begin{equation}
    \rho_d= \left(\frac{B}{1+A-\sqrt{3}\zeta_0}+ \frac{C_1}{a^{3(1+\alpha)(1+A-\sqrt{3}\zeta_0)}} \right)^{\frac{1}{1+\alpha}}
\end{equation}
where $C_1$ is an integrating constant. The above expression can
be further re-written as:
\begin{equation}
    \rho_d= \rho_{d0}\{A_s+ (1-A_s)(1+z)^{3(1+\alpha)(1+A-\sqrt{3}\zeta_0)} \}^{\frac{1}{1+\alpha}}
\end{equation}
where $\rho_{d0}$ being the DE density value at the
present epoch, $A_s= \frac{B}{(1+A-\sqrt{3}\zeta_0)C_1+B}$
satisfying the conditions $0<A_s<1$ and $1+A-\sqrt{3}\zeta_0 >0$,
and $\rho_{d0}^{1+\alpha}=
\frac{(1+A-\sqrt{3}\zeta_0)C_1+B}{1+A-\sqrt{3}\zeta_0}$. Thus, we
can write the Hubble parameter from equation (\ref{frw-1}) as
follows:
\begin{multline} \label{hubble_viscous-MCG}
    H^2(z)=H_0^2 \bigg [\Omega_{m0}(1+z)^3 +\Omega_{d0} \Big \{A_s+ (1-A_s)(1+z)^{3(1+\alpha)(1+A-\sqrt{3}\zeta_0)} \Big \}^\frac{1}{1+\alpha}\bigg] \times \\
   \bigg [1-\frac{3H_0^2}{\rho_1} \Big \{\Omega_{m0}(1+z)^3 +\Omega_{d0} \Big\{A_s+ (1-A_s)(1+z)^{3(1+\alpha)(1+A-\sqrt{3}\zeta_0)} \Big \}^\frac{1}{1+\alpha}\Big \} \bigg]
\end{multline}
where $\Omega_{m0}=\frac{\rho_{m0}}{3H_0^2}$ and
$\Omega_{d0}=\frac{\rho_{d0}}{3H_0^2}$ are the dimensionless
parameters all defined at the present epoch. Further, substituting
$z=0$ in equation (\ref{hubble_viscous-MCG}) and equating the RHS
to `$1$' we get:
\begin{equation}
    \Big [\Omega_{m0} +\Omega_{d0} \Big] \times \bigg [1-\frac{3H_0^2}{\rho_1} \Big \{\Omega_{m0}+\Omega_{d0}\Big \} \bigg]=1
\end{equation}
which simplifies to
\begin{equation}\label{Od0}
\Omega_{d0}=\frac{1}{6H_0^2}\left(\rho_1+\sqrt{\rho_1^2-12\rho_1
H_0^2 } \right) -\Omega_{m0}
\end{equation}
\subsection{Variable modified Chaplygin gas (VMCG)}
The expression for pressure in the case of variable modified
Chaplygin gas (VMCG) is given by \cite{41}:
\begin{equation}
    p_d= C\rho_d -\frac{D(a)}{\rho_d^\beta}
\end{equation}
where $0 \leq \beta \leq 1$ and $C>0$. Assuming, $D(a)=D_0 a^{-n}$
where $D_0 >0$ and $n>0$, we get from Eq (\ref{de}):
\begin{equation}
    \rho_d= \left[\frac{3(1+\beta)D_0}{\{3(1+\beta)(1+C)-n\}a^n} + \frac{K}{a^{3(1+C)(1+\beta)}} \right]^{\frac{1}{1+\beta}}
\end{equation}
where $K(>0)$ is an arbitrary integrating constant and
$3(1+\beta)(1+C)-n >0$. Also, here $n$ must be positive, or else
for $a \rightarrow \infty$ we would have $\rho_d \rightarrow
\infty$ thereby contradicting our case for expanding Universe. The
expression for $\rho_d$ above can be further simplified to:
\begin{equation}
    \rho_d =\rho_{d0} \left[\frac{C_s}{a^n}+ \frac{1-C_s}{a^{3(1+C)(1+\beta)}} \right]^{\frac{1}{1+\beta}}
\end{equation}
where $\rho_{d0}$ denotes the DE density value at the present
epoch, $C_s= 1-\frac{K}{\rho_{d0}^{1+\beta}}$ and
$\rho_{d0}^{1+\beta}= K +
\frac{3(1+\beta)D_0}{3(1+\beta)(1+C)-n}$. Thus, we can write the
Hubble parameter from equation (\ref{frw-1}) as follows:
\begin{multline}\label{hubble_variable-MCG}
    H^2(z)=H_0^2 \bigg [\Omega_{m0}(1+z)^3 +\Omega_{d0} \Big \{C_s(1+z)^n + (1-C_s)(1+z)^{3(1+C)(1+\beta)} \Big \}^{\frac{1}{1+\beta}} \bigg] \times \\
   \bigg [1-\frac{3H_0^2}{\rho_1} \Big \{\Omega_{m0}(1+z)^3 +\Omega_{d0} \Big\{C_s(1+z)^n + (1-C_s)(1+z)^{3(1+C)(1+\beta)} \Big \}^{\frac{1}{1+\beta}} \Big \} \bigg]
\end{multline}
where $\Omega_{d0}$ is given in (\ref{Od0}).\\
\section{Methodology}\label{methodology}
In our study, we carefully picked a specific set of recent Baryon Acoustic Oscillation (BAO) measurements from various galaxy surveys, with the primary contributions coming from observations made by the Sloan Digital Sky Survey (SDSS) \cite{42,43,44,45,46,47}. We also included valuable data from the Dark Energy Survey (DES) \cite{48} , the Dark Energy Camera Legacy Survey (DECaLS) \cite{49}, and 6dFGS BAO \cite{50} to enhance the diversity of our dataset. Additionally, we incorporated thirty uncorrelated Hubble parameter measurements obtained through the cosmic chronometers (CC) method discussed in \cite{50,51,52,53}. Furthermore, we included the latest Pantheon sample data on Type Ia Supernovae \cite{54}, 24 binned quasar distance modulus data from \cite{55}, a set of 162 Gamma-Ray Bursts (GRBs) as outlined in \cite{56}, and the recent Hubble constant measurement (R22) \cite{57} as an additional prior. In our analysis, we employed a nested sampling approach implemented in the open-source Polychord package \cite{58}, complemented by the GetDist package \cite{59} to present our results in a clear and informative manner. We also utilized Markov Chain Monte Carlo (MCMC) to obtain the values of free parameters of both proposed cosmological models and conduct the likelihood analysis.\\\\
\begin{figure}[htbp]
\centering
\includegraphics[scale=0.58]{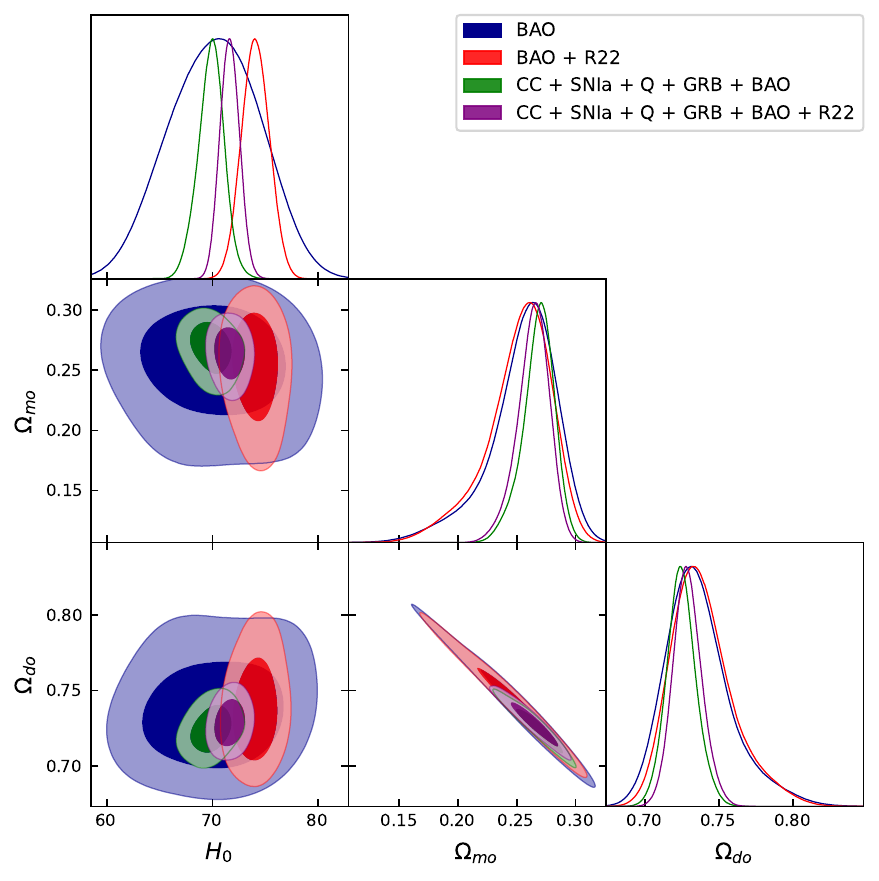}
\caption{The figure illustrates 1$\sigma$ and 2$\sigma$ regions using the standard $\Lambda$CDM model to demonstrate the posterior distribution of various observational data measurements.}\label{fig_lcdm}
\end{figure}
\begin{figure*}[htbp]
\centering
\includegraphics[scale=0.4]{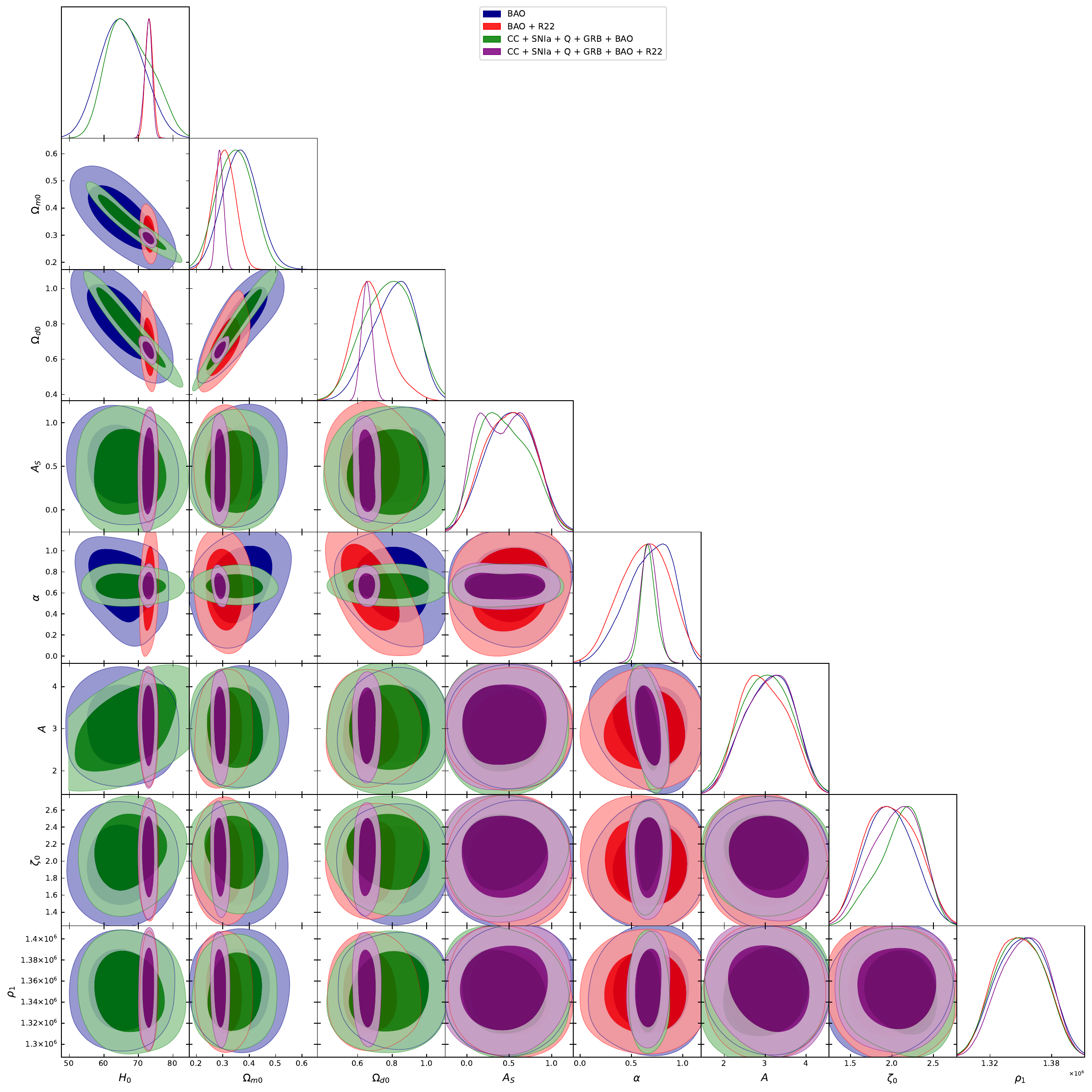}
\caption{The figure illustrates the posterior distribution of various observational data measurements using the Viscous Modified
Chaplygin Gas model, highlighting the 1$\sigma$ and 2$\sigma$ regions.}\label{fig_2}
\end{figure*}
\begin{figure*}[htbp]
\centering
\includegraphics[scale=0.4]{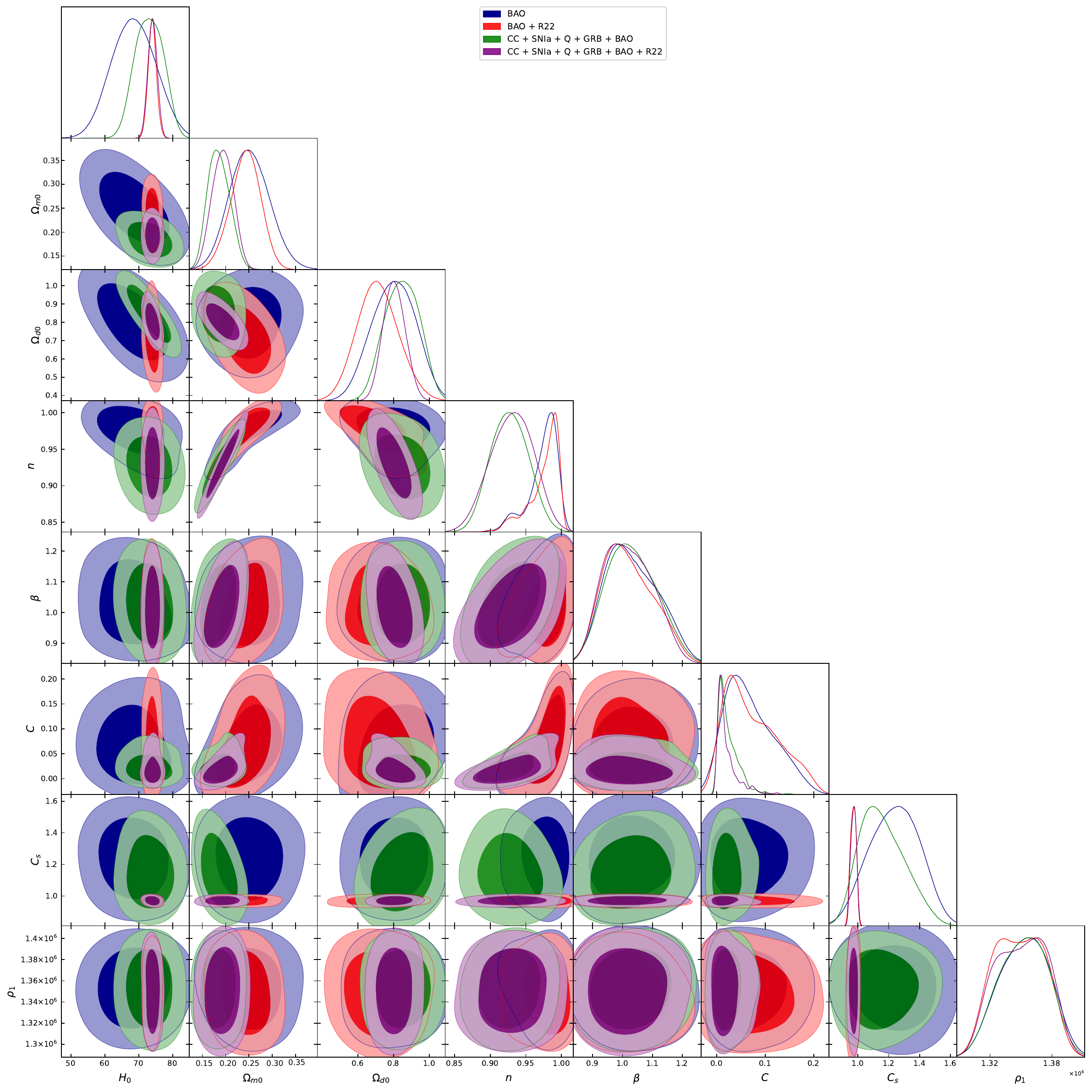}
\caption{The figure illustrates the posterior distribution of various observational data measurements using the Variable Modified
Chaplygin Gas model, highlighting the 1$\sigma$ and 2$\sigma$ regions.}\label{fig_3}
\end{figure*}
\begin{sidewaystable}
\centering
\begin{tabular}{|c|c|c|c|c|c|c|}
\hline
\multicolumn{7}{|c|}{MCMC Results} \\
\hline\hline
Model & Parameters & Priors & BAO & BA0 + R22 & CC + SC + BAO & CC + SC + BAO + R22 \\[1ex]
\hline
& $H_0$ & [50,100] & $69.089209_{\pm 4.396874}^{\pm 6.252595}$ & $73.906596_{\pm 1.357697}^{\pm 2.788642}$ & $69.854848_{\pm 1.259100}^{\pm 2.386935}$ & $71.616475_{\pm 1.003940}^{\pm 1.936192}$ \\[1ex]
$\Lambda$CDM Model &$\Omega_{m0}$ &[0.,1.]  & $0.256796_{\pm 0.025772}^{\pm 0.069423}$ & $0.254859_{\pm 0.026912}^{\pm 0.068716}$ & $0.268654_{\pm 0.012822}^{\pm 0.028134}$ & $0.264407_{\pm 0.013179}^{\pm 0.030224}$ \\[1ex]
&$\Omega_{d0}$  & [0.,1.] & $0.734631_{\pm 0.020888}^{\pm 0.038211}$ & $0.736400_{\pm 0.021510}^{\pm 0.034343}$ & $0.724585_{\pm 0.009373}^{\pm 0.016999}$ & $0.728562_{\pm 0.009532}^{\pm 0.018197}$ \\[1ex]
& $r_d$ (Mpc) & [100,200] & $149.504236_{\pm 10.037166}^{\pm 15.212831}$ & $139.447508_{\pm 2.914639}^{\pm 5.883881}$ & $146.543556_{\pm 2.598566}^{\pm 5.101856}$ & $143.299915_{\pm 2.218062}^{\pm 4.358875}$ \\[1ex]
& $r_{d}/r_{fid}$ & [0.9,1.1] & $1.008382_{\pm 0.066344}^{\pm 0.101188}$ & $0.940599_{\pm 0.020806}^{\pm 0.037308}$ & $0.990266_{\pm 0.019826}^{\pm 0.035810}$ & $0.967284_{\pm 0.015101}^{\pm 0.030444}$ \\
\hline
& $H_0$ & [50,100] &$ 65.168094^{\pm 6.617797}_{\pm 10.434466}$ & $73.07201^{\pm 0.978415}_{\pm 1.889797}$ & $ 67.447204  ^{\pm 2.514723}_{\pm 1.878032}$ & $72.860428^{\pm 1.040048}_{\pm 2.328246}$ \\[1ex]
&$\Omega_{m0}$ &[0.,1.]  &$ 0.365902 ^{\pm 0.069461}_{\pm 0.115990}$ & $ 0.305203 ^{\pm 0.041952}_{\pm 0.071096}$ & $ 0.346451 ^{\pm 0.069036}_{\pm 0.105164}$ & $ 0.289643 ^{\pm 0.012301}_{\pm 0.025106}$   \\[1ex]
VsMCG Model &$\Omega_{d0}$ &[0.,1.] &$0.809753^{\pm 0.133984}_{\pm 0.234411}$  & $0.676112^{\pm 0.101291}_{\pm 0.190216}$ & $0.779877^{\pm 0.158516}_{\pm 0.239042}$ & $0.652845^{\pm 0.028308}_{\pm 0.059635}$ \\[1ex]
&$A_{s}$  & [0.,1.] &$0.514312_{\pm 0.279102}^{\pm 0.491146}$  & $0.504348_{\pm  0.324263}^{\pm 0.457075} $ & $0.436312 _{\pm 0.319504 }^{\pm 0.416373}$ & $0.447568 _{\pm 0.357146}^{\pm 0.430062}$ \\[1ex]
&$\alpha$  & [0.,1.] &$0.683658_{\pm 0.232932}^{\pm 0.445186}$ &$0.628384_{\pm 0.262042}^{\pm 0.440275}$ & $0.665943_{\pm 0.064040}^{\pm 0.144111}$ & $0.667478_{\pm 0.074594}^{\pm 0.157306}$ \\[1ex]
&$A$  & [2.,4.] &$3.089772_{\pm 0.649737}^{\pm 1.049832}$ & $2.965524 _{\pm 0.602096}^{\pm 0.891859}$ &  $3.014177_{\pm 0.699794}^{\pm 0.952994}$ & $3.093859 _{\pm 0.615597}^{\pm 1.043732}$ \\[1ex]
&$\zeta_{0}$  & [1.5,2.5] &$1.962968_{\pm 0.310779}^{\pm 0.442674}$ & $1.990460 _{\pm 0.349834}^{\pm 0.453422}$ &  $2.101817_{\pm 0.297352}^{\pm 0.555898}$ & $2.050346 _{\pm 0.357694}^{\pm 0.511089}$ \\[1ex]
&$\rho_{1}$  & [1310000,1390000] &$1351709.4_{\pm 28642.8}^{\pm 38561.9}$ & $1349170.2 _{\pm 26320.8}^{\pm 36390.3}$ &  $1349393.1_{\pm 26568.3}^{\pm 37491.8}$ & $1353509.1 _{-25630.2}^{+41653.0}$ \\[1ex]
& $r_{d}$ (Mpc) & [100,200] & $149.571115_{\pm 8.752594}^{\pm 16.426449}$ & $140.757158_{\pm 8.454083}^{\pm 12.992500}$ & $147.559044_{\pm 2.723020}^{\pm 5.811905}$ & $141.158148_{\pm 2.299581}^{\pm 4.709648}$ \\[1ex]
& $r_{d}/r_{fid}$ & [0.9,1.1] &$1.009424_{\pm 0.059881}^{\pm 0.104352}$ & $0.990016_{\pm 0.056765}^{\pm 0.083577}$ & $0.995630_{\pm 0.019954}^{\pm 0.039894}$ & $0.993363_{\pm 0.016175}^{\pm 0.031652}$ \\[1ex]
\hline
& $H_0$ & [50,100] & $65.796665^{\pm 6.461974}_{\pm 10.070874}$ & $73.799085^{\pm 1.325015}_{\pm 2.574862}$ & $69.204717^{\pm 2.413623}_{\pm 1.056031}$ & $73.025435^{\pm 1.390057}_{\pm 2.671485}$ \\[1ex]
&$\Omega_{m0}$ &[0.,1.] &$0.286515^{\pm 0.075623}_{\pm 0.172756}$&$0.197851^{\pm 0.052402}_{\pm 0.100870}$&$0.229029^{\pm 0.065811}_{\pm 0.145501}$&$0.181731^{\pm 0.060864}_{\pm 0.119488}$   \\[1ex]
VMCG Model &$\Omega_{d0}$ &[0.,1.] &$0.700150^{\pm 0.113069}_{\pm 0.186913}$ & $0.803863^{\pm 0.129291}_{\pm 0.276353}$ & $0.801456^{\pm 0.120813}_{\pm 0.212190}$ & $0.717771^{\pm 0.067980}_{\pm 0.116377}$ \\[1ex]
&$n$  & [0.5,1.5] & $0.990322^{\pm 0.007463}_{\pm 0.034757}$ & $0.980549^{\pm 0.017220}_{\pm 0.048783}$ & $0.976961^{\pm 0.018299}_{\pm 0.051484}$ & $0.970191^{\pm 0.026968}_{\pm 0.065479}$\\[1ex]
&$\beta$ & [0.8,1.4] &$1.018741^{\pm 0.086082}_{\pm 0.115003}$ & $1.020759^{\pm 0.084889}_{\pm 0.112896}$ & $1.021356^{\pm 0.085708}_{\pm 0.115778}$ & $1.025479^{\pm 0.097275}_{\pm 0.123108}$\\[1ex]
& $C$ & [0.,0.3] & $0.082870^{\pm 0.057201}_{\pm 0.079413}$& $0.082870^{\pm 0.057201}_{\pm 0.079413}$& $0.045065^{\pm 0.036463}_{\pm 0.044343}$ & $0.038790^{\pm 0.033014}_{\pm 0.037825}$\\[1ex]
&$C_{s}$  & [0.,1.] & $0.599738^{\pm 0.361244}_{\pm 0.545885}$ & $0.599738^{\pm 0.361244}_{\pm 0.545885}$ & $0.376236^{\pm 0.293035}_{\pm 0.364827}$ & $0.481279^{\pm 0.380702}_{\pm 0.472094}$ \\[1ex]
&$\rho_{1}$  & [1310000,1390000] & $1351317.6^{\pm 27841.6}_{\pm 39991.2}$ & $1349811.1^{\pm 25642.0}_{\pm 36678.6}$ & $1354908.4^{\pm 26738.7}_{\pm 42104.6}$ & $1350850.4^{\pm 27481.4}_{\pm 37505.8}$ \\[1ex]
& $r_{d}$ (Mpc) &[100,200]& $148.683108_{\pm 8.949488}^{\pm 14.573363}$ & $139.365379_{\pm 9.762761}^{\pm 14.205048}$ & $147.906003_{\pm 2.386071}^{\pm 4.444459}$ & $141.623321_{\pm 2.638046}^{\pm 5.305695}$ \\[1ex]
& $r_{d}/r_{fid}$ &[0.9,1.1] & $1.002078_{\pm 0.063968}^{\pm 0.097008}$ & $1.000626_{\pm 0.067223}^{\pm 0.094807}$ & $0.995166_{\pm 0.016201}^{\pm 0.030789}$ & $0.994314_{\pm 0.019600}^{\pm 0.035566}$ \\[1ex]
\hline
\end{tabular}
\caption{The table presents constraints on cosmological parameters for the $\Lambda$CDM, VsMCG, and VMCG models at a 95\% confidence level.}\label{tab_2}
\end{sidewaystable}

\begin{figure*}[htb]
\begin{subfigure}{.32\textwidth}
\includegraphics[width=\linewidth]{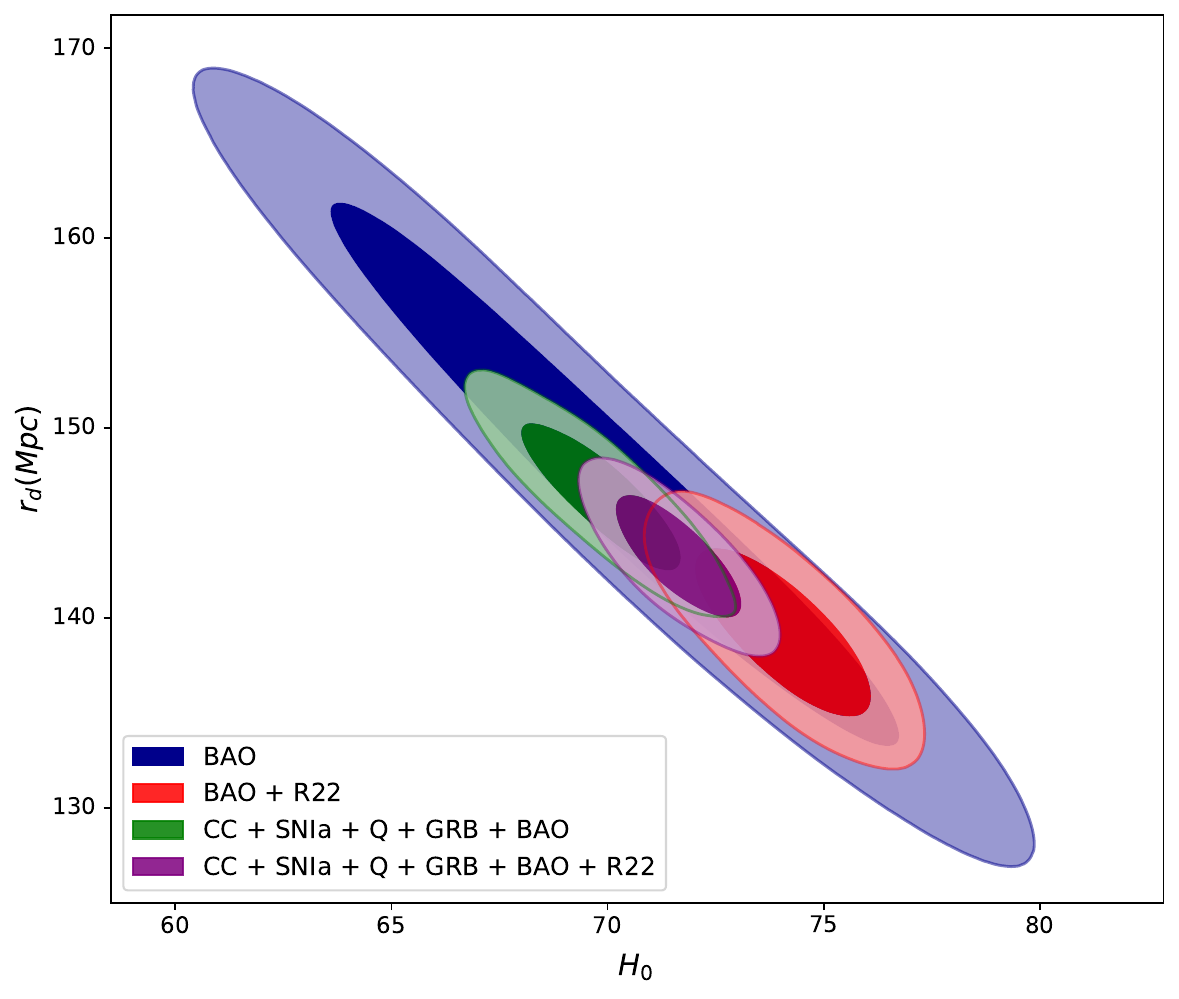}
    \caption{$\Lambda$CDM Model}
    \label{fig_4a}
\end{subfigure}
\hfil
\begin{subfigure}{.32\textwidth}
\includegraphics[width=\linewidth]{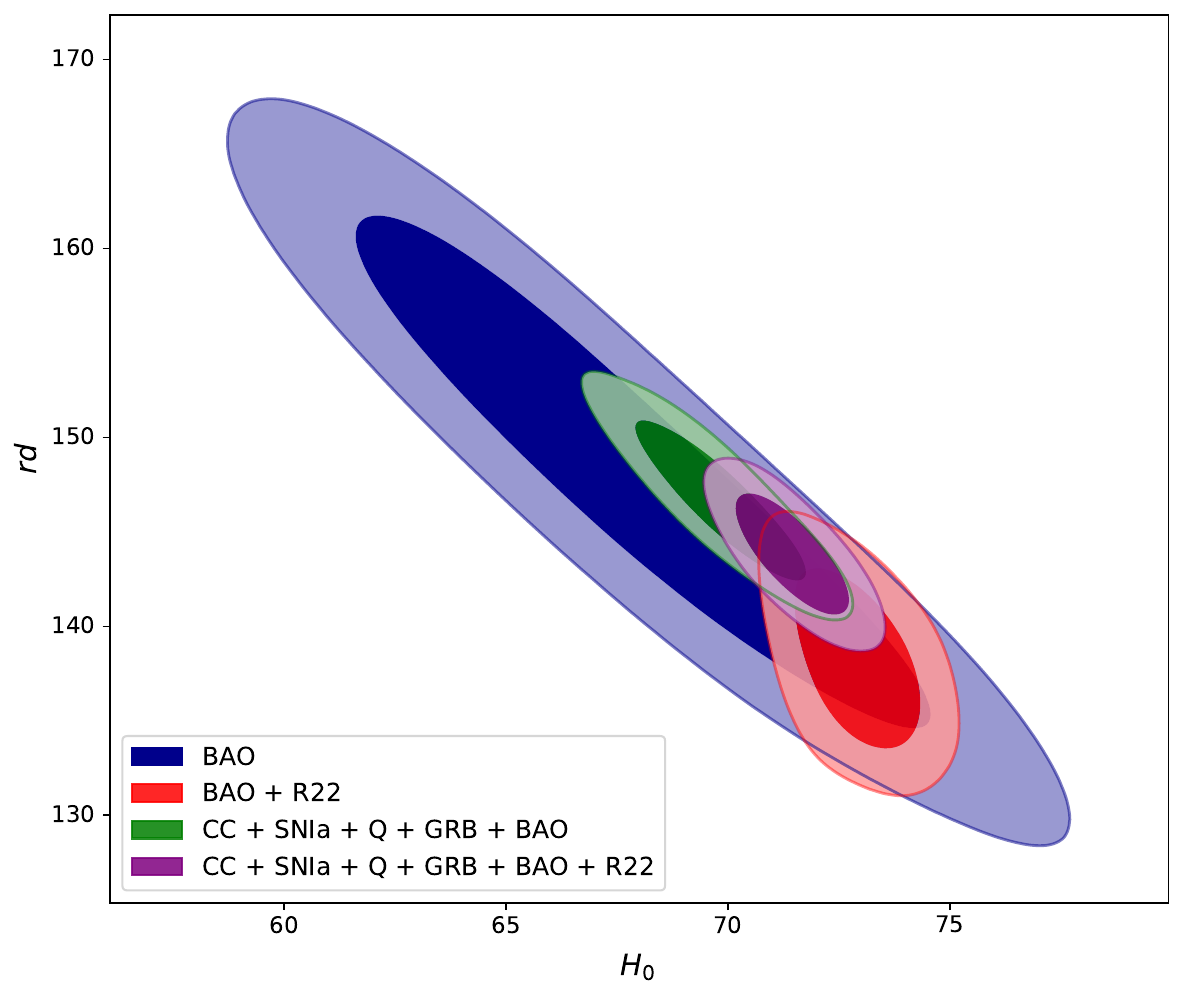}
    \caption{VsMCG Model}
    \label{fig_4b}
\end{subfigure}
\hfil
\begin{subfigure}{.32\textwidth}
\includegraphics[width=\linewidth]{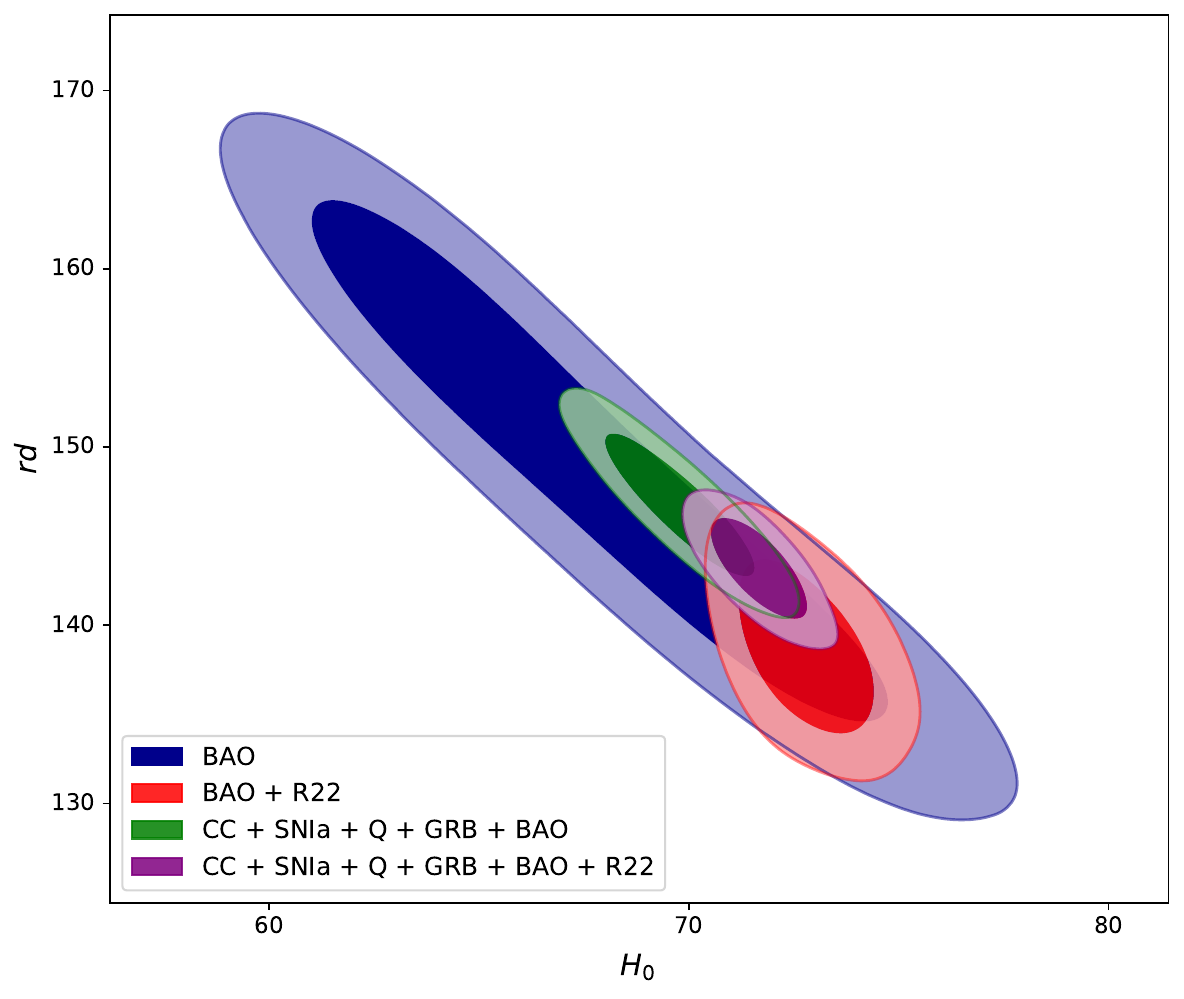}
    \caption{VMCG Model}
     \label{fig_4c}
\end{subfigure}
\caption{The figure shows the posterior distribution of
diverse observational data measurements within the \(r_{d}-H_{0}\)
contour plane using the $\Lambda$CDM, VsMCG and VMCG models. The
shaded regions correspond to the 1$\sigma$ and 2$\sigma$
confidence plane.}
\end{figure*}

\section{Gravitational Lensing} \label{lensing}
The differential probability that a distribution of
galaxies per unit redshift will be multiply imaged
by a source with redshift $z_s$ is \cite{38}:

\begin{equation} \label{OD-1}
    \frac{d \tau}{d z_l}= n(\Delta \theta,z_l) (1+z_l)^3 S \frac{c dt}{d z_l}
\end{equation}
where $n(\Delta \theta,z_l)$ denotes the co-moving number density of lenses, $S$ is the
lens cross-section and $c dt/d z_l$ the proper distance
interval. Further, following numerous work \cite{34,60} done
previously, we assume an singular isothermal sphere (SIS) profile
for our lensing model. Thus, the expression for lensing
cross-section in an SIS profile is given by:
\begin{equation}
   S=16\pi^3 \left(\frac{\sigma}{c}\right)^4 \left(\frac{D_l D_{ls}}{D_s}\right)^2
\end{equation}
where $\sigma$ is the velocity dispersion, $D_s$, $D_{l}$ and
$D_{ls}$ represents the angular diameter distances between observer-source,
observer-lens and lens-source respectively, given by:
\begin{equation}
    D_{z_1z_2} =\frac{1}{\sqrt{H_0}(1+z_2)}\int_{z_1}^{z_2} \frac{1}{\sqrt{H(z)}}dz \hspace{1cm} (k=0)
\end{equation}
and finally the proper distance interval reads:
\begin{equation}
    \frac{cdt}{dz_l}= \frac{c}{(1+z_l)H(z_l)}
\end{equation}
Now, to calculate the velocity dispersion function $\sigma$, we use the relation \cite{61}:
\begin{equation} \label{modified-Schechter}
     \frac{dn}{d\sigma}=n_* \left(\frac{\sigma}{\sigma_*} \right)^P e^{-(\sigma/\sigma_*)^Q} \frac{Q}{(P/Q)} \frac{1}{\sigma}
\end{equation}
where $P$ is the faint slope, $Q$ the high velocity cut-off and, $n_*$ and $\sigma_*$ the characteristics number density and characteristics velocity dispersion respectively. Further, we assume redshift evolution of the parameters as follows: $n_*(z_l) =n_*(1+z_l)^\gamma $ and $\sigma_*(z_l) = \sigma_*(1+z_l)^\nu $
where $\gamma$ and $\nu$ are free parameters.
Also, following the works of Choi et al.
\cite{62} to derive the
velocity dispersion function of early type
galaxies, we have $n_*=8 \times 10^{-3} h^3 Mpc^{-3}$ where
$h$ is $H_0$ in units of $100 km s^{-1} Mpc^{-1}$, $\sigma_*= 161 km s^{-1}$, $P=2.32 \pm 0.10$, and $Q=2.67 \pm 0.07$\\\\
Therefore, from equations (\ref{OD-1}) and (\ref{modified-Schechter}), we deduce a straightforward analytical expression for the optical depth of a point source at redshift $z_s$ with image separation $\Delta \theta$ in an FRW Universe with dark matter and dark energy components \cite{61,38}:
    \begin{equation} \label{optical-depth}
    \frac{d \tau}{d z_l}(\Delta \theta, z_l)= \tau_N (1+z_l)^{(\gamma- \nu P)} (1+z_l)^3 \frac{D_{ls}}{D_s} D_{l}^2 \frac{cdt}{dz_l} \left(\frac{\Delta \theta}{\Delta \theta_*} \right)^{1+P/2} \text{exp}\left[-\left(\frac{\Delta \theta}{\Delta \theta_*} \right)^{Q/2} (1+z_l)^{-\nu Q} \right]
\end{equation}
where $\tau_N= 2\Pi^2 n_* \left(\frac{\sigma_*}{c} \right) \frac{Q}{\Gamma(P/Q)}$ is the normalization factor and $\Delta \theta_*= 8 \pi \left(\frac{\sigma_*}{c} \right) \frac{D_{ls}}{D_s}$.
\begin{figure}
    \centering
    \includegraphics[keepaspectratio=true,scale=0.5]{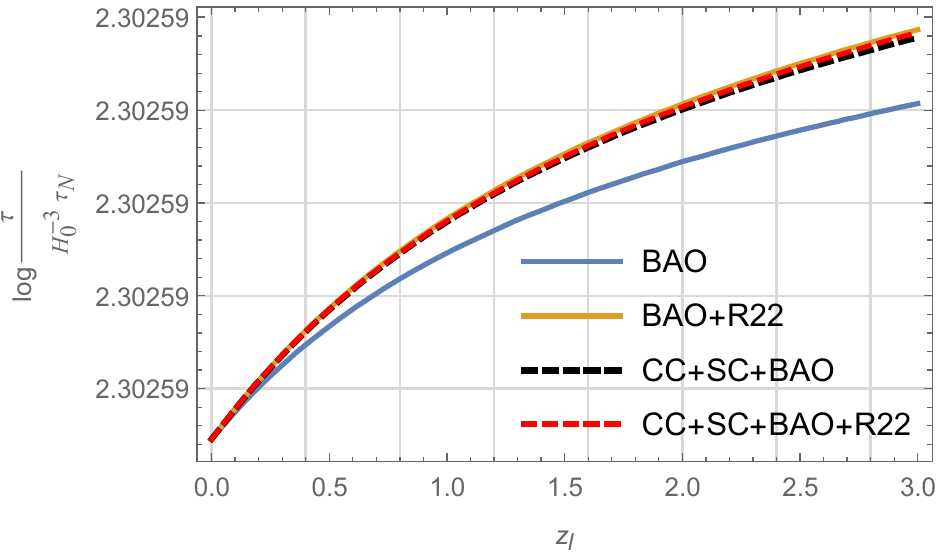}
    \caption{Variation in optical depth behaviour of VsMCG model w.r.t $z_l$ for parameters constrained through different data sets given in Table \ref{tab_2}.}
    \label{fig1}
\end{figure}
\begin{figure}
    \centering
    \includegraphics[keepaspectratio=true,scale=0.5]{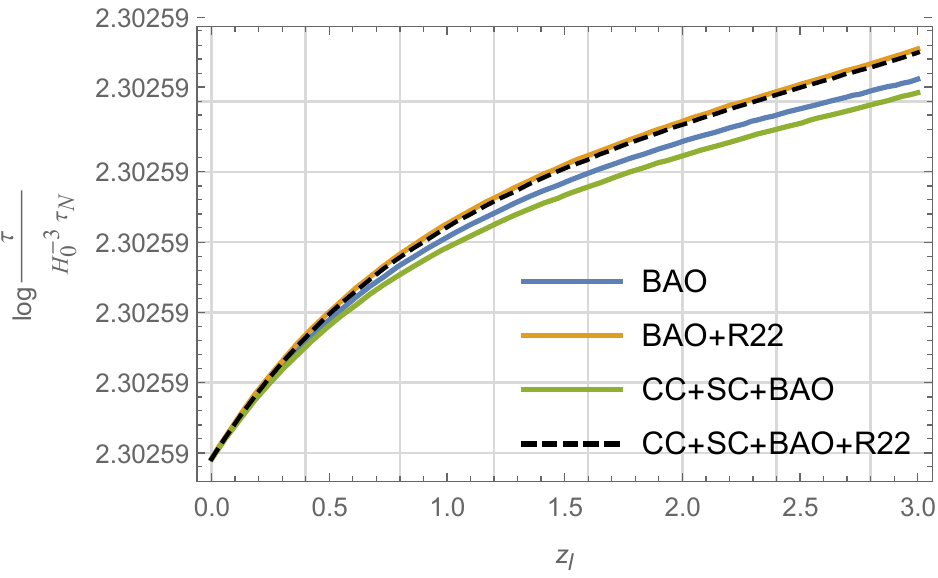}
    \caption{Variation in optical depth behaviour of VMCG model w.r.t $z_l$ for parameters constrained through different data sets given in Table \ref{tab_2}.}
    \label{fig2}
\end{figure}
\begin{figure}
    \centering
    \includegraphics[keepaspectratio=true,scale=0.5]{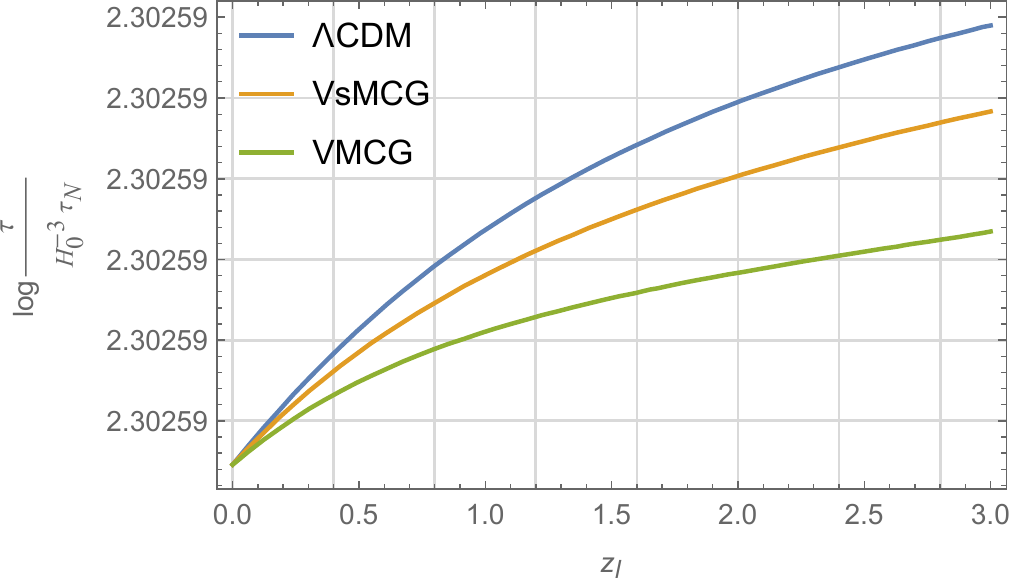}
    \caption{Variation in optical depth behaviour of $\Lambda$CDM, VsMCG and VMCG model w.r.t lens redshift $z_l$ for parameter values given in Table \ref{tab_2}.}
    \label{fig3}
\end{figure}
\section{Results} \label{results}
For the standard $\Lambda$CDM model, the posterior distribution of the key cosmological parameters is shown at the \(68\%\) and \(95\%\) confidence levels in Fig. \ref{fig_lcdm}.
Detailed results from MCMC simulations are outlined in Table \ref{tab_2}. When the R22 prior is incorporated into the joint dataset, the optimal fitting value for \(H_{0}\) is \(71.50 \pm 0.882\), diverging \cite{20} from but aligning closely with measurements from the SNIe sample in \cite{57}. In the absence of R22 priors with the Joint dataset, the estimated \(H_{0} = 69.837 \pm 1.204\) corresponds more closely to the value in \cite{20}. Figs \ref{fig_2} and \ref{fig_3} portray the \(68\%\) and \(95\%\) confidence levels for crucial cosmological parameters in VsMCG and VMCG models. Across these models, when the R22 prior is included in the Joint dataset, the optimal fitting value for \(H_{0}\) diverges from \cite{20} but aligns more closely with the SNIe sample in \cite{57}. Conversely, without R22 priors and using the Joint dataset, the estimated \(H_{0} = 69.837 \pm 1.204\) aligns more closely with \cite{20}. The determined values for the matter density, denoted as $\Omega_{m0}$, and the dark energy density, denoted as $\Omega_{d0}$, in the $\Lambda$CDM and VMCG Models, appear to be lower compared to the values documented in \cite{20} ($\Omega_{m0}=0.315 \pm 0.007$, $\Omega_{d0}=0.685 \pm 0.007$). However, in the case of the VsMCG Model, it is higher than the value documented in \cite{20}. However, this observation has been documented in alternative studies \cite{63,64}. In the context of the Baryon Acoustic Oscillations (BAO) scale, it is defined by the cosmic sound horizon imprinted in the cosmic microwave background during the drag epoch, denoted as \(z_d\). This epoch marks the separation of baryons and photons. The BAO scale, represented by \(r_d\), is determined by the integral of the ratio of the speed of sound (\(c_s\)) to the Hubble parameter (\(H_{0}\)) over the redshift range from \(z_d\) to infinity. The speed of sound, \(c_s\), is given by \(\sqrt{\frac{\delta p_\gamma}{\delta \rho_B + \delta \rho_\gamma}}\), where \(\delta p_\gamma\) is the pressure perturbation in photons, and \(\delta \rho_B\) and \(\delta \rho_\gamma\) are perturbations in the baryon and photon energy densities, respectively. This expression is further simplified to \(\frac{1}{\sqrt{3(1+R)}}\), with \(R\) defined as the ratio of baryon density perturbation to photon density perturbation (\(R \equiv \frac{\delta \rho_B}{\delta \rho_\gamma} = \frac{3 \rho_B}{4 \rho_\gamma}\)). Observational data from \cite{20} provides the redshift at the drag epoch as \(z_d = 1059.94 \pm 0.30\). In the case of a $\Lambda$CDM model, measurements from \cite{20} estimate the BAO scale \(r_d\) to be \(147.09 \pm 0.26\) megaparsecs (Mpc). In our exploration of the $\Lambda$CDM model, Fig \ref{fig_4a} illustrates the posterior distribution of the contour plane for \(r_{d}-H_{0}\). The BAO datasets provide an estimated \(r_d\) of \(145.807 \pm 9.06\) Mpc, aligning with the reported findings in \cite{65}. However, when we exclusively incorporate R22 prior into the BAO dataset, the determined sound horizon at the drag epoch is \(138.345 \pm 2.45\) Mpc. Examining the Joint dataset reveals an estimated BAO scale (\(r_d\)) of \(145.811 \pm 2.34\) Mpc, closely aligning with the outcomes reported in \cite{20}. Additionally, incorporating the R22 prior into the comprehensive dataset results in \(r_d\) of \(142.591 \pm 1.49\) Mpc, indicating proximity to the findings in \cite{66}. Now, shifting to the VsMCG Model, Fig \ref{fig_4b} presents the posterior distribution for the \(r_{d}-H_{0}\) contour plane. For the BAO datasets, the resulting \(r_{d}\) is \(149.571 \pm 8.752\) Mpc, consistent with the Planck results as reported in \cite{65}. However, incorporating R22 prior exclusively into the BAO dataset leads to a sound horizon at the drag epoch of \(140.757 \pm 8.454\) Mpc. In the case of the Joint dataset, the determined \(r_{d}\) is \(147.559 \pm 2.72\) Mpc, aligning closely with the Planck results. Furthermore, integrating the R22 prior into the full dataset results in \(r_{d} = 141.158 \pm 2.299\) Mpc, showing proximity to the findings presented in \cite{66}. In the context of the VMCG Model, Fig \ref{fig_4c} illustrates the posterior distribution of the \(r_{d}-H_{0}\) contour plane. When considering BAO datasets, the estimated BAO scale (\(r_{d}\)) is \(148.683 \pm 8.949\) Mpc, a result consistent with the findings reported in \cite{65}. However, introducing the R22 prior into the BAO dataset alone yields a sound horizon at the drag epoch of \(139.365 \pm 9.762\) Mpc. In the case of the Joint dataset, the calculated BAO scale (\(r_{d}\)) is \(147.906 \pm 2.386\) Mpc, which closely matches the results obtained from the Planck mission. Additionally, when we include the R22 prior in the complete dataset, the resulting \(r_{d}\) is \(141.623 \pm 2.638\) Mpc, indicating a close agreement with the findings reported in \cite{66}. The results from $\Lambda$CDM, VsMCG and VMCG models exhibit tension with the \(r_{d}\) value estimated by Planck. However, these results still demonstrate agreement with the findings presented in \cite{66} reports that by employing Binning and Gaussian methods to combine measurements of 2D BAO and SNIa data, the values of the absolute $\mathrm{BAO}$ scale range from 141.45 Mpc to $r_{d} \leq 159.44 \mathrm{Mpc}$ (Binning) and 143.35 Mpc to $r_{d} \leq 161.59 \mathrm{Mpc}$ (Gaussian). These findings highlight a clear discrepancy between early and late-time observational measurements, analogous to the $H_{0}$ tension. It is noteworthy that our results are contingent on the range of priors for $r_{d}$ and $H_{0}$, influencing the estimated values in the $r_{d}-H_{0}$ contour plane. An interesting observation is that when we exclude the R22 prior, the results for $H_{0}$ and $r_{d}$ tend to align with the Planck and SDSS results, mitigating the tension observed in the absence of this particular prior. Now, in the study of the gravitational lensing optical depth, we
use the analytical expression given by eqn. \eqref{optical-depth}.
After constraining the model parameters of the DE models as
discussed above and finding their best-fit values, we plot the
graphs of $log(\tau(z_l)/H_0^{-3} \tau_N)$ against redshift $z_l$ for both 
models. Figs. \ref{fig1} and \ref{fig2}, respectively, show changes
in the optical depth behaviour of the two DE models here w.r.t $z_l$.
We notice that the probability of finding gravitational lenses
(i.e. optical depth) increases for both the models with increasing
$z_l$ value. Further, from fig. \ref{fig1}, we observe that the
graphs of $log(\tau(z_l)/H_0^{-3} \tau_N)$ vs $z_l$ for parameter 
sets constrained through
different data sets show similar behaviour except for the
parameters constraint through $BAO$ only. However, for $z_l <0.3$,
they all coincide with each other. Again, if we analyse fig.
\ref{fig2}, we observe that parameters constrained through the
dataset of $CC+SC+BAO$ gives a lower probability of finding
gravitational lenses as compared to the other three combinations.
Also, combinations of the dataset $BAO+R22$ and $CC+SC+BAO+R22$
show same lensing probability for this model (VMCG). From these, 
we can conclude that
the parameter values constraint through different data sets have
very little to no effect in detecting the distribution of
gravitational lenses in our Universe.
Finally, we conclude our study
by plotting the optical depth of the two DE models considered here
along with $\Lambda$CDM against lens redshift $z_l$. We observe
from fig. \ref{fig3} that although at the present epoch, the
gravitational lensing probability for the three models coincides
with each other, however moving back in time (or higher redshift
values), all the models highly diverge from each other with
$\Lambda$CDM giving the highest lensing probability and VMCG the
lowest.
\section{Discussions and Conclusions} \label{conclusion}
Assuming our Universe to be filled with dark matter and dark
energy combinations in a flat FRW Universe, we reviewed two
models of modified Chaplygin gas as the fluid source. For the dark
energy candidates, we considered viscous modified Chaplygin gas
(VsMCG) and variable modified Chaplygin gas (VMCG) in the
framework of loop quantum cosmology. We determined the respective
Hubble parameter and, consequently, constraint the associated
model parameters with the help of the latest observational data. Our investigation involved the selection of 24 Baryon Acoustic Oscillation (BAO) points, 30 uncorrelated data points from Cosmic Chronometers, 40 points from Type Ia supernovae, 24 points from the Hubble diagram for quasars, and 162 points from Gamma Ray Bursts. Additionally, we integrated the most recent
measurement of the Hubble constant conducted by researcher R22.
Our recent investigation, conducted through a comprehensive array
of observational assessments, brings attention to the persistent
existence of the Hubble tension, although it is somewhat mitigated
to a $2 \sigma$ level for $H_0$. By introducing the sound horizon
($r_d$) as a free parameter, we have derived specific values for
$H_0$ and $r_d$ in various cosmological models, including the
Standard $\Lambda$CDM, VsMCG, and VMCG models. In the $\Lambda$CDM
model, our analysis reveals $H_0 = 69.828145 \pm 1.009964 \
\mathrm{km/s/Mpc}$ and $r_d = 146.826212 \pm 2.201612 \
\mathrm{Mpc}$. For the VsMCG model, we obtain $H_0 = 67.447204 \pm
8.878032 \ \mathrm{km/s/Mpc}$ and $r_d = 147.559044 \pm 2.723020 \
\mathrm{Mpc}$. Similarly, in the case of VMCG, the results
indicate $H_0 = 69.204717 \pm 8.056031 \ \mathrm{km/s/Mpc}$ and
$r_d = 147.906003 \pm 2.386071  \ \mathrm{Mpc}$. Crucially, our
assessments highlight that the values of $H_0$ and $r_d$ based on
low-redshift measurements align with early Planck
estimates~\cite{1BAO}. We also studied the gravitational lensing
optical depth of our two dark energy models with the help of the
constrained model parameter values. The graph showing changes in
optical depth is plotted against their lens redshift (fig.
\ref{fig1} \& \ref{fig2}). For both models, the probability of
finding gravitational lenses increases with higher redshift
values. However, for low redshift, the lensing
probability in both the DE models coincides irrespective of the
different sets of constrained parameter values obtained by
considering different combinations of data sets. Lastly, we also
carried out a joint analysis of the two DE models considered here
with that of $\Lambda$CDM model (fig. \ref{fig3}). We found that
the three models are indistinguishable for low redshift value
($z_s<1)$; however, they highly diverge from each other in the
early Universe. Finally, we conclude our study with a statistical
evaluation of our cosmological models and utilize both the Akaike
Information Criterion (AIC) and the Bayesian Information Criterion
(BIC). The AIC is expressed as \cite{67,68}: $\mathrm{AIC} =
-2 \ln(\mathcal{L}_{\text{max}}) + 2k +
\frac{2k(2k+1)}{N_{\text{tot}} - k - 1}$ Here,
\(\mathcal{L}_{\max}\) signifies the maximum likelihood of the
data, incorporating the entire dataset without the R22 prior. The
parameters encompass \(N_{\text{tot}}\), the total number of data
points (in our instance, \(N_{\text{tot}} = 296\)), and \(k\), the
number of parameters. For large \(N_{\text{tot}}\), this
expression simplifies to:$\mathrm{AIC} \simeq -2
\ln(\mathcal{L}_{\max}) + 2k$ which is the conventional form of
the AIC criterion \cite{67}. In contrast, the Bayesian
Information Criterion is formulated as \cite{72BAO}: $\mathrm{BIC}
= -2 \ln(\mathcal{L}_{\max}) + k \ln N_{\text{tot}}$. By applying
these criteria, we compute the AIC and BIC for the standard
\(\Lambda\)CDM, VsMCG, and VMCG models. The obtained values for
\(\Lambda\)CDM, VsMCG, and VMCG models are, respectively AIC =
$[277.38, 281.16, 281.12]$ and BIC = $[277.59, 283.12, 282.34]$.
Despite the \(\Lambda\)CDM model displaying the best fit due to
the lowest AIC, our collective AIC and BIC results lend support to
all the tested models. This suggests that none of the models can
be dismissed based on the existing data. In the evaluation of the
VsMCG and VMCG models relative to $\Lambda$CDM, we acknowledge
that $\Lambda$CDM is embedded within both proposed extensions,
differing by 5 degrees of freedom. This distinction allows for the
application of standard statistical tests. The yardstick for
comparison is the reduced chi-square statistic, defined as
$\chi_{\text{red}}^{2} = \chi^{2} / \text{Dof}$, where Dof
represents the degrees of freedom of the model, and $\chi^{2}$
denotes the weighted sum of squared deviations with an equal
number of runs for the three models, the statistic approximates 1,
expressed as: $ \left(\frac{\chi^{2}}{Dof_{\Lambda CDM}},
\frac{\chi^{2}}{Dof_{VsMCG}}, \frac{\chi^{2}}{Dof_{VMCG}} \right)
\approx \{0.961646, 0.971234, 0.954310\}$. This comparative
analysis offers valuable insights into the goodness of fit for
each model, with values near 1 signifying a satisfactory alignment
with the observed data. The findings suggest that both the VsMCG
and VMCG models provide viable alternatives to the standard
$\Lambda$CDM model, with comparable goodness of fit and support
from statistical criteria.

\end{document}